\documentclass[aps,prl,preprint,superscriptaddress]{revtex4-2}
\usepackage{amsmath, bm}
\usepackage{graphicx}
\usepackage{subfigure}  % use for side-by-side figures
\usepackage{textgreek}  % for muon symbol outside of math mode
\usepackage[sort&compress]{natbib} % for references range

\begin{document}

%\preprint{}

\title{Information and decoherence in a muon-fluorine coupled system}


\author{J. M. Wilkinson}
\email[]{john.wilkinson@physics.ox.ac.uk}
%\thanks{}
\affiliation{Clarendon Laboratory, University of Oxford Department of Physics, Parks Road, Oxford, OX1 3PJ, United Kingdom}

\author{S. J. Blundell}
\email[]{stephen.blundell@physics.ox.ac.uk}
%\thanks{}
\affiliation{Clarendon Laboratory, University of Oxford Department of Physics, Parks Road, Oxford, OX1 3PJ, United Kingdom}


\date{\today}

\begin{abstract}
  {\bf The unitary evolution of a quantum
  system preserves its coherence, but interactions between the system
  and its environment result in decoherence, a process in which the
  quantum information stored in the system becomes degraded.  A
  spin-polarized positively-charged muon implanted in a fluoride
  crystal realises such a coherent quantum system, and the entanglement
  of muon and nearest-neighbour fluorine nuclear spins gives rise to
  an oscillatory time-dependence of the muon polarization which 
  can be detected and measured.  Here we show that the decohering
  effect of more distant nuclear spins can be precisely modelled, allowing a
  very detailed description of the decoherence processes coupling
  the muon-fluorine `system' with its `environment', and allowing 
  us to track the system entropy as the quantum information degrades.
  These results show how to precisely quantify the spin relaxation of 
  muons implanted into quantum entangled states in fluoride crystals, a 
  feature that has hitherto only been described phenomenologically.  }
\end{abstract}

%\keywords{}

\maketitle

%\section{Introduction}
An important issue in the study of quantum mechanics is the
interaction between a {\it system}, $\mathcal{S}$, considered as a few coupled
quantum objects evolving in a manner described by some well-defined
Hamiltonian, and its {\it environment}, $\mathcal{E}$, considered as a large bath
consisting of many quantum objects.  The action of the environment is
to act as a source of decoherence \cite{zurek} whereby quantum
information, stored in the system and in principle readable from it,
is degraded and leaks out into the environment where it can no longer
be discovered.  If the system and environment could be considered
together as a single system, $\mathcal{S}\otimes\mathcal{E}$,
this larger system would undergo unitary evolution and its von 
Neumann entropy, $S=-\mbox{Tr}\rho\log_2\rho$, where $\rho$ 
is the density matrix of the $\mathcal{S}\otimes\mathcal{E}$ composite 
object, would be constant.  However, we are rarely permitted this 
holistic view and are restricted to monitoring the reduced density 
matrix of the system, $\rho_{\mathcal{S}} = \mbox{Tr}_{\mathcal{E}}\rho$, 
obtained by tracing out the degrees of freedom of the
environment \cite{joos}, and the entropy of $\mathcal{S}$ will tend to increase with time
\cite{vedral}.

\begin{figure}[hp]
\includegraphics[width=\textwidth]{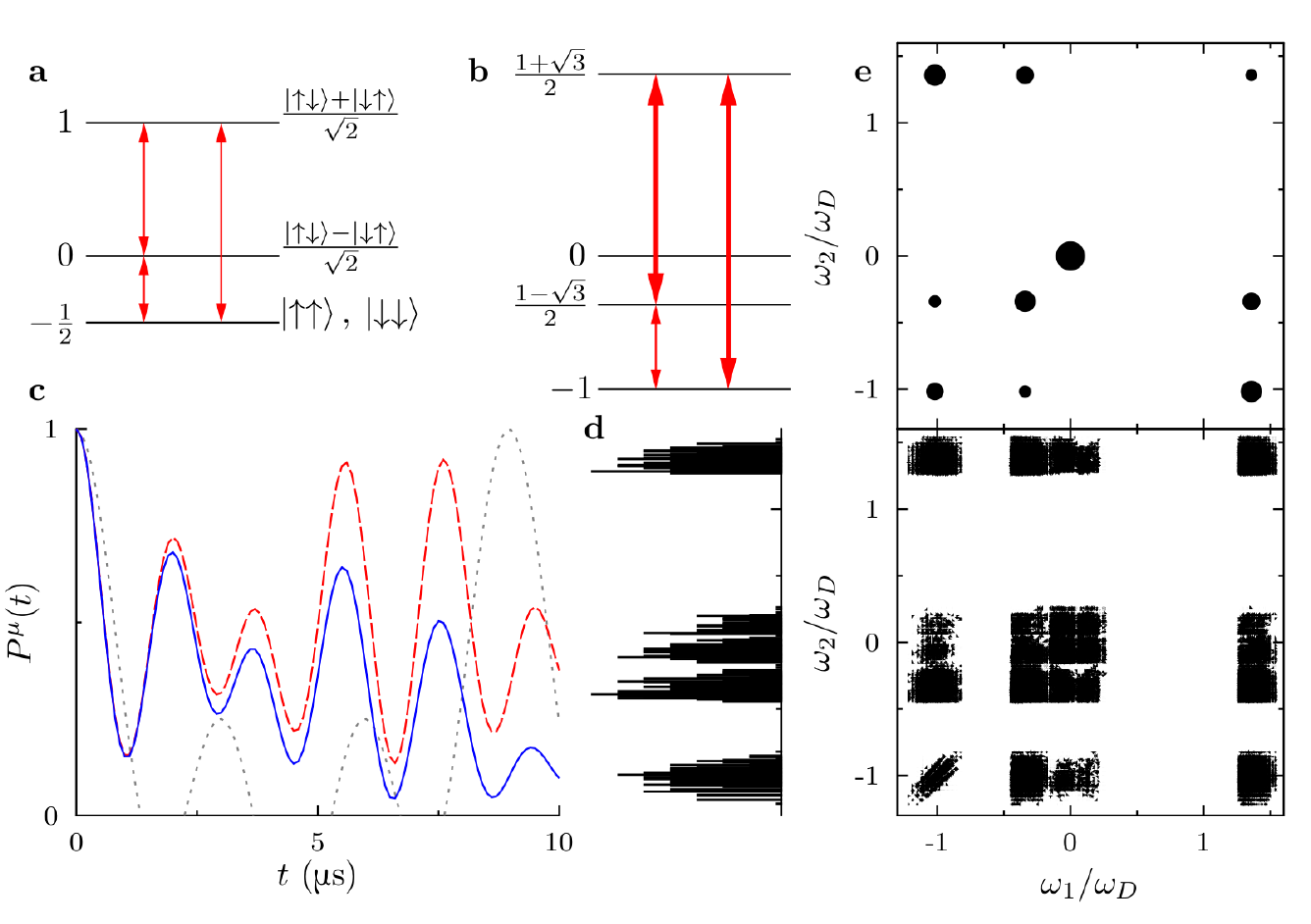}
\caption{{\bf Muon-fluorine coupled states.}
  {\bf a}, The energy levels in a F--$\mu$ coupled state.  The
  eigenstates are labelled by the spins of the muon and the fluorine nucleus.
  The red arrows show possible transitions and the energies are in units
  of $\hbar\omega_{\rm D} = \hbar\mu_0\gamma_\mu\gamma_F/(4\pi r^3)$.
  {\bf b}, The energy levels for a F--$\mu$--F state, also showing
  possible transitions (the dominant transitions are shown with
  thicker lines).  The energy eigenvalues are very slightly different from
  those shown once the small F--F dipolar coupling is included, as will be
  done in all subsequent plots.
  {\bf c}, The time-dependence of the muon polarization $P^\mu(t)$ for isolated
  F--$\mu$ (dotted line), isolated F--$\mu$--F (dashed line) and for
  F--$\mu$--F also coupled to eight next-nearest-neighbour fluorine
  nuclear spins appropriate for the fluorite structure. These simulations
  are for an experiment in zero applied magnetic field, and assume a 
  polycrystalline average over all possible orientations of the F--$\mu$ or
  F--$\mu$--F species.
  {\bf d}, Energy levels for the F--$\mu$--F state including
  next-nearest-neighbour fluorine nuclear spins.
  {\bf e}, Transition diagram for isolated F--$\mu$--F (top panel) and
  with the next-nearest-neighbour couplings (bottom panel). The strength
  of the couplings between the $\hbar\omega_1$ and $\hbar\omega_2$
  is represented by the relative areas of the points. The  energy scales of 
  {\bf b} and {\bf d} are lined up with these two panels.
}
\label{fig:Fmu}
\end{figure}

In order to study decoherence experimentally, it is necessary to identify
well-defined scenarios in which the interaction between the system and
environment is well characterised.  One such scenario is provided by the
interaction between a spin-polarized positively-charged muon $\mu^+$
and the neighbouring nuclei in a fluoride compound.  Fluorine nuclei
have spin $I=\frac{1}{2}$ with 100\% abundance, and fluoride ions are
very electronegative, making their surroundings attractive sites for 
$\mu^+$.  Often a F--$\mu$--F
species forms after muon implantation, resulting in a distinctive
oscillatory signal measured in the positron decay asymmetry
\cite{brewer}, a direct result of the entanglement between the
fluorine and muon spins \cite{lancaster}.  The dipolar interaction 
between a single fluorine nuclear spin and a muon would result in 
the energy level spectrum shown in Fig.~\ref{fig:Fmu}a, while for 
two fluorine nuclear spins (the F--$\mu$--F state), the spectrum 
is shown in Fig.~\ref{fig:Fmu}b. In both cases, the distinctive beating 
pattern of oscillations in the time-dependence of the muon polarization $P^\mu_z(t)$
occurs because of transitions between these energy levels. This effect 
can be interpreted as a coherent exchange of spin polarization between 
the initially polarized muon and the initially unpolarized fluorine nuclei. 
These oscillations are shown in Fig.~\ref{fig:Fmu}c and have been
observed in numerous inorganic fluorides
\cite{brewer,noakes,cs2agf4}, fluoropolymers
\cite{pratt,nishiyama,teflon} and fluoride-containing molecular
magnets \cite{lancaster}.  However, in all cases good fits to the experimental data
have only been obtained by multiplying the calculated coherent
precession signals by a phenomenological relaxation function, often a
stretched exponential, the parameters of which have no theoretical
basis. A master equation approach could be used to model the
non-unitary evolution of the reduced density matrix of the system
\cite{lindblad}, but this would still involve an arbitrary parameter
quantifying the system-environment coupling.  We will show below that
an exact treatment is possible which includes the known couplings
between the muon and
more distant fluorine nuclei, thereby accurately modelling the {\it
  environment} of the F--$\mu$--F {\it system}. These couplings result in a relaxation
of the precession signal (solid line in Fig.~\ref{fig:Fmu}c) that
completely accounts for the data and makes contact with recent
electronic structure calculations of the muon site \cite{moller,bernardini}.

The effects of interactions with more distant fluorine nuclei can be
understood by examining the energy eigenvalues shown in
Fig.~\ref{fig:Fmu}d where the eight next-nearest neighbour (nnn) couplings
in the fluorite structure have also been included.  The
four energy levels in isolated F--$\mu$--F are broadened by the nnn couplings into four bands of
energy levels.  The transitions between these energy levels are shown
in the two-dimensional plots in Fig.~\ref{fig:Fmu}e, where the size of the point indicates the
strength of the transition from energy levels $\hbar\omega_1$ to
$\hbar\omega_2$. These diagrams are reminiscent of two-dimensional NMR plots \cite{aue}, but here there are no radiofrequency pulses and the transitions happen automatically in the unitary evolution of the quantum state. Thus, the overall structure of the transitions for isolated F--$\mu$--F in the upper panel is
largely retained in the lower panel when including the more distant couplings, but a
richer frequency spectrum results and this mixture of frequencies is
responsible for the dephasing of the precession signal observed in
experiments.

\begin{figure}[hp]
\includegraphics[width=\textwidth]{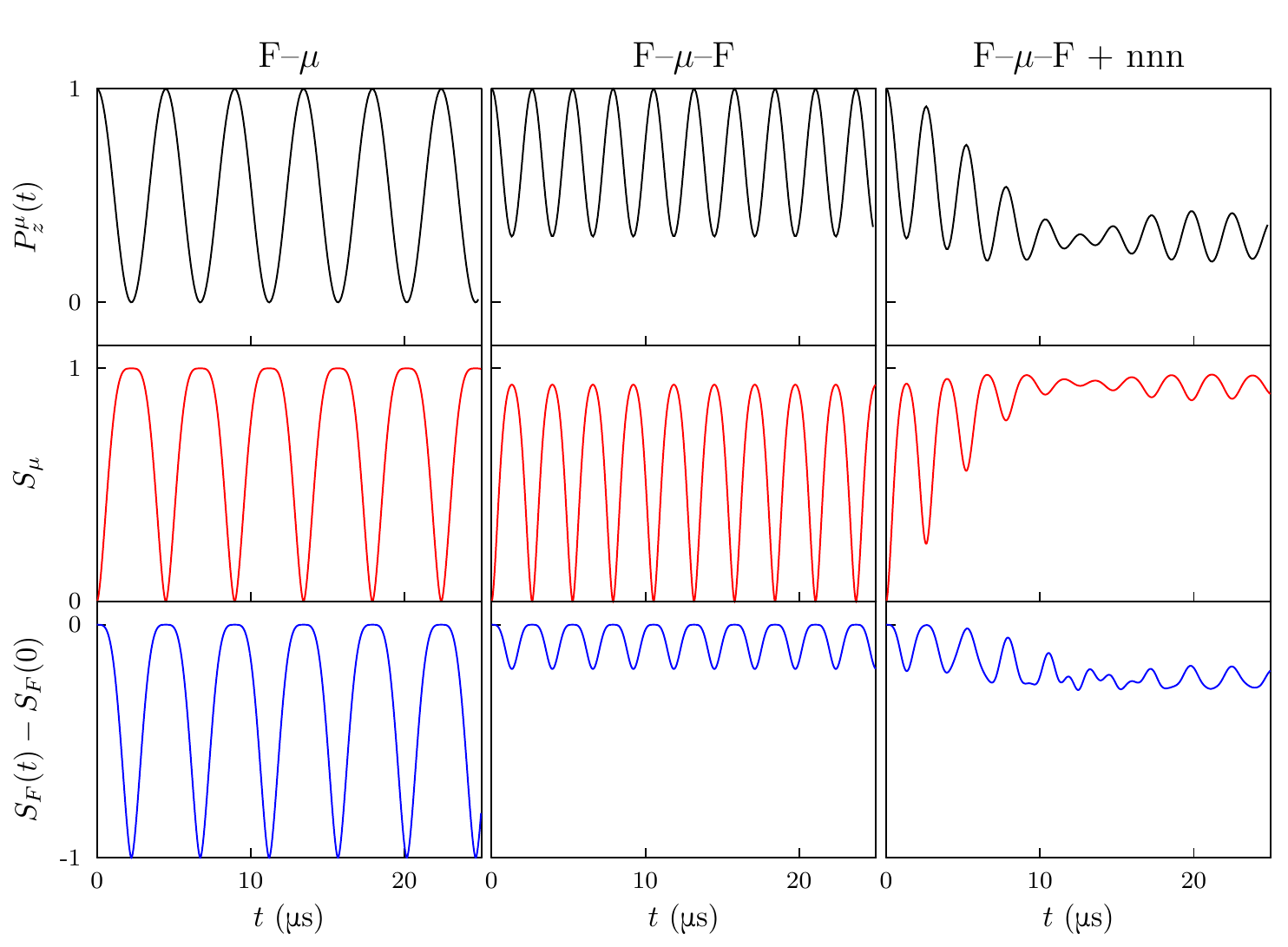}
\caption{{\bf von Neumann entropy for muon-fluorine states.}
The time-dependence of the muon polarization $P_z^{\mu}(t)$, the muon
entropy $S_\mu$ (obtained by tracing out all the other spins), and the
entropy of the entire fluorine system, $S_F$ (note that $S_F(0)=N_F$, 
and our von Neumann entropies use $\log_2$, so that information
is measured in bits. These are plotted for the three cases of isolated 
F--$\mu$, isolated F--$\mu$--F and environmentally decohering F--$\mu$--F. 
These simulations assume the F--$\mu$ (or F--$\mu$--F) bond is aligned 
with the initial muon spin polarization. (The other case is treated in 
Supplementary Fig. 1, and shows similar behaviour.) 
}
\label{fig:S}
\end{figure}

Further insight can be gained by calculating the time-dependence of
the von Neumann entropy. We consider three cases: (i) the F--$\mu$ 
state; (ii) the F--$\mu$--F state; and (iii) the F--$\mu$--F state with 
eight nnn fluorine nuclei, appropriate for the fluorite structure \cite{hayes}. The von 
Neumann entropy for these states remains constant at $S=N_F$
as the states evolve unitarily, where $N_F$ is the number of fluorine
nuclei in the cluster ($N_F=1,~2$ and $10$ for the three cases, respectively).
This is because the implanted muon is initially spin-polarized and hence
in a pure state, but the fluorine nuclei are initially unpolarized. By tracing
out the fluorine or muon degrees of freedom, we are able to calculate the 
muon and fluorine reduced entropies, $S_\mu$ and $S_F$, as a function of time,
see Fig.~\ref{fig:S}. The coupling between the muon and its fluoride environment
results in the muon oscillating between being in a completely pure ($S_\mu=0$)
and mixed ($S_\mu>0$) state, with the fluorine subsystem oscillating in antiphase.
This can be interpreted in terms of quantum information exchanging back and
forth between the muon and the fluorine subsystem; $P_z^\mu(t)$ reaches a 
maximum whenever information is stored on the muon and a minimum 
whenever it is residing in the fluoride subsystem. For F--$\mu$, there are times 
when the muon is in a completely mixed state and the fluorine nucleus is in a
completely pure state, but for F--$\mu$--F the fluoride subsystem never
evolves into a pure state. However, for both F--$\mu$ and F--$\mu$--F, 
the muon periodically returns to a completely pure state ($S_\mu=0$) 
and the quantum information is therefore never lost. 

However, when the effect of the eight additional nnn fluorines is included, the
muon never recovers to a pure state within the timescale of a typical muon experiment 
(25 \textmu s). Thus the eight nnn fluorines act as a source of decoherence, 
so that information transferred from the muon remains in this subsystem and 
never completely returns to the muon. This results in the oscillations in $P_z^\mu(t)$
exhibiting relaxation. However, even including nnn interactions only results in
a larger interacting cluster and does not yet account for the decoherence due 
to the entire crystal, an issue we will return to.

To demonstrate how to account for system-environment interactions,
we identified CaF$_2$ as a model system since the Ca nuclear spin
can be neglected (the most abundant Ca isotopes have $I=0$ and make up
99.86\% of the nuclei); thus only the fluorine nuclei contribute to the
$\mu$SR spectrum.  CaF\textsubscript{2} adopts the cubic fluorite structure 
(lattice parameter $a = 5.451$ \AA), and the muon site has been identified 
by density functional theory calculations (DFT$+\mu$ \cite{moller}, see Methods). 
The muon site lies between two fluoride ions, each of which is pulled in towards the
muon, resulting in a 14\% reduction in the F--F separation distance.
These calculations show that the effect of the muon on the positions of 
the more distant nuclei is negligible. We used an exact diagonalization method to 
evaluate the time evolution of the density matrix and simulate $P_z^\mu(t)$
(see Methods), rather than one of the approximate techniques that are sometimes 
employed \cite{celio}. This has the virtue of accounting for all interactions precisely, 
but the dimension of the Hilbert space is $2\prod^M_{i=1}(2I_i+1)$, where the product is over the $M$
nuclei included in the calculation, and this grows exponentially with $M$, making this method 
prohibitively computationally expensive when too many nuclei are included. 
Hence we restrict our diagonalization method to include only nearest-neighbour 
and nnn fluorine nuclei, but scale the nnn interactions to account for all couplings
in the infinite lattice. This can be done in a quantitative way by considering the 
second moment of the nuclear dipole field distribution, a quantity well known 
from the theory of broadening of NMR lines \cite{vanvleck,abragam}. The second 
moment $\sigma_M^2$ of this distribution at the muon site is given by
\begin{equation}
\sigma_M^2 = \frac{2}{3}\Big(\frac{\mu_0}{4\pi}\Big)^2\hbar^2\gamma_\mu^2\sum_{j=1}^M\frac{\gamma_j^2I_j(I_j+1)}{r_j^6},
\end{equation}
where $r_j$ is the distance from the muon to the $j$\textsuperscript{th} nucleus with
spin $I_j$ and gyromagnetic ratio $\gamma_j$, $\gamma_\mu $($=2\pi\times135.5$ MHz T$^{-1}$)
is the muon gyromagnetic ratio, and the sum converges as $M\to\infty$.
 We then calculate $\lambda$ such that 
\begin{equation}
\sigma_\infty^2 = \sigma_{\rm{nn}}^2 + \frac{2}{3}\Big(\frac{\mu_0}{4\pi}\Big)^2\hbar^2\gamma_\mu^2\sum_{j\in \rm{nnn}}\frac{\gamma_j^2I_j(I_j+1)}{(\lambda r_j)^6},
\label{eq:nnnbroadening}
\end{equation}
where $\sigma_{\rm{nn}}^2$ is due to nearest neighbour couplings only and the 
sum is restricted to nnn. Thus we adjust our coupling to the nnn nuclei using the 
parameter $\lambda$ to mimic the effect of all more distant couplings. Because 
contributions to the second moment scale as $1/r_j^6$, we expect $\lambda$ 
to be close to unity (but $\lambda<1$ because the more distant couplings make 
a positive contribution to $\sigma_\infty^2$). Completing this calculation for the 
case of CaF$_2$, we find that $\lambda = 0.937$ (see Methods) 

\begin{figure}[hp]
  \includegraphics[width=0.9\textwidth]{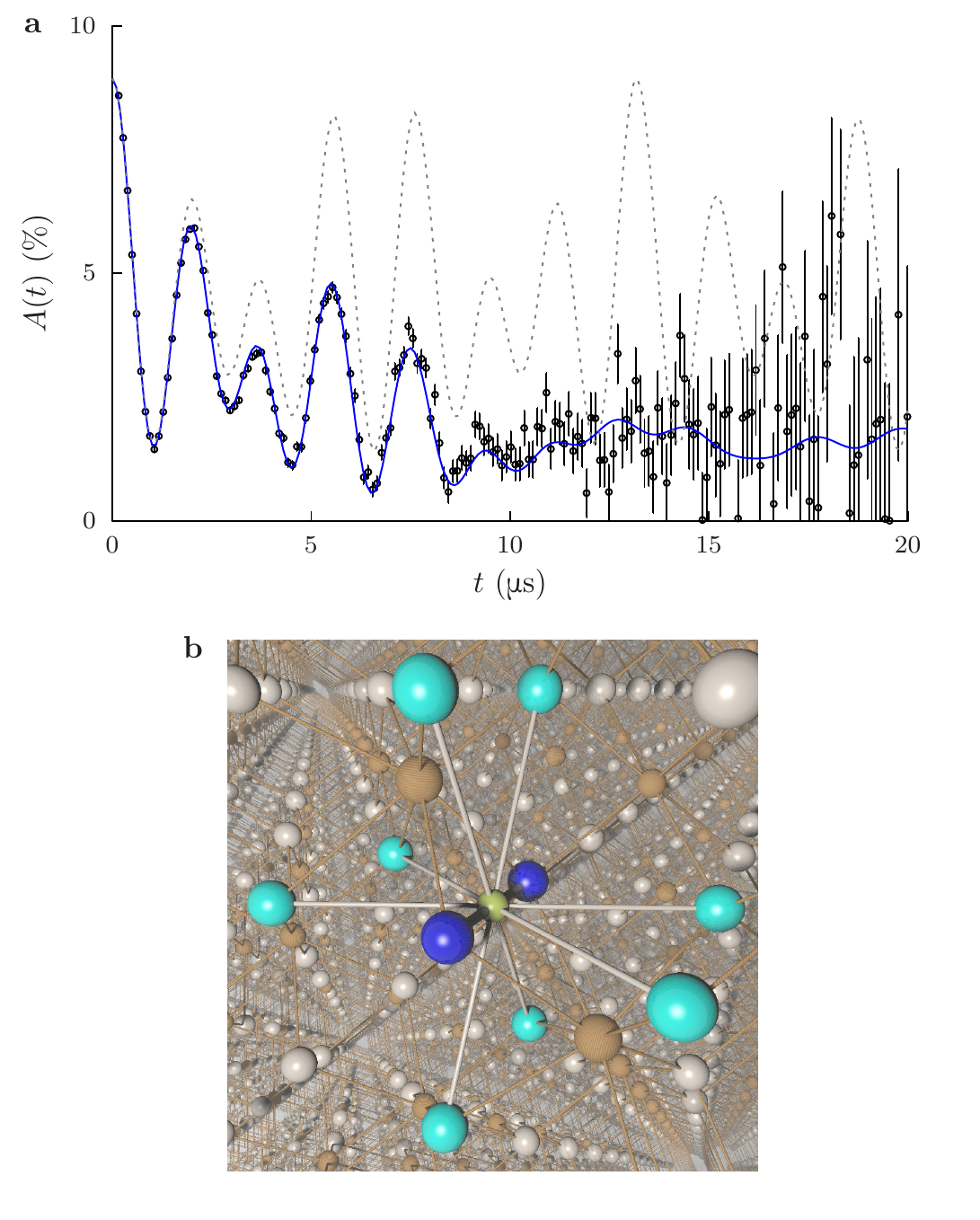}
\caption{{\bf Muon-fluorine decoherence in CaF$_2$.}
  {\bf a}, Muon decay asymmetry data $A(t)$ for polycrystalline 
  CaF$_2$, together with the simulated muon
  polarization without (dotted line) and with (solid line) the effects
  of environmental decoherence. The error bars on the data are calculated
  by considering the number of muon decays measured at the corresponding
  time.
  {\bf b}, The muon (yellow sphere) strongly coupled to two fluorine
  nuclei (dark blue spheres), and weakly coupled to next-nearest
  neighbour fluorine nuclei (cyan spheres), embedded inside the
  fluorite structure of CaF$_2$.
}
\label{fig:CaF2}
\end{figure}

The agreement of these simulations with the experimentally observed
$A(t)$ can be seen in Fig.~\ref{fig:CaF2}a. 
If only the nearest-neighbour fluorine nuclei are considered (isolated
F--$\mu$--F, dashed line in Fig.~\ref{fig:CaF2}a) the fit is very poor, but the 
inclusion of nnn couplings results in an impressive agreement between
theory and experiment (solid line in Fig.~\ref{fig:CaF2}a). Note that 
this fit does not need to include any phenomenological relaxation function of 
the sort used in all previous studies \cite{brewer,lancaster,noakes,cs2agf4,pratt,nishiyama,teflon}.
Instead, the observed relaxation of the oscillations results entirely from the nnn couplings.
Our fit uses only two fitting parameters, one of which is the distance between the
muon and the two nearest-neighbour fluorine nuclei, which is found to be 
$1.172(1)$ \AA~(very close to the DFT prediction of $1.134$ \AA, and dramatically
shorter than the $\frac{a}{4}=1.362$ \AA~expected if there was no muon-induced 
distortion). The second fitting parameter is $\lambda=0.920(3)$, within 2\% of our 
predicted value. These results demonstrate that, with suitable scaling, the
eight nnn fluoride ions, which constitute a spin-subspace of dimensionality $2^8=256$,
can provide an adequate representation of the full environment due to the entire crystal
(Fig.~\ref{fig:CaF2}b), allowing a quantitative description of the decoherence for this problem.

We also now demonstrate that this method can be extended to the more general case, in 
which the cation nuclear spin is non-negligible. For example, NaF adopts the rocksalt
structure and contains sodium nuclei which have a spin of $I=\frac{3}{2}$. In this case,
the muon forms an F--$\mu$--F state with the two nearest fluorine nuclei, but the next
largest couplings arise from the sodium nuclei. In this case, we tried using the two sodium
nuclei (subspace dimension 16) as a proxy for all more distant fluoride and sodium couplings,
and evaluated the muon polarization function only for these five spins (one muon, two fluorines, 
and two sodiums, with dimensionality 128). This proved sufficient to account for the measured
relaxation and gave parameters consistent with our DFT+$\mu$ calculations (see Supplementary 
Information).

In summary, we have found that the couplings between fluorine nuclei and positive muons can act as an ideal 
model system to observe the effects of quantum information dissipation through decoherence. We expect
our method to find wide applicability in quantitatively describing decohering relaxation in experiments
on a wide range of other crystalline materials.

\newpage
\section*{Methods}

\subsection*{$\mu$SR experiments}

In the muon experiment, a beam of spin-polarized muons were
incident on a sample, and the number of positrons detected in both
the forwards and backwards detectors, $N_F(t)$ and $N_B(t)$
respectively was measured \cite{cox}. The muon asymmetry was
calculated as
\begin{equation}
A(t) = \frac{N_B(t)-\alpha N_F(t)}{N_B(t) + \alpha N_F(t)},
\label{eq:asymmetry}
\end{equation}
where the parameter $\alpha$ takes into account systematic differences
between the readings of both sets of detectors.  Our experiments were
performed using the MuSR spectrometer at the ISIS Facility, Rutherford
Appleton Laboratory, UK.  A polycrystalline sample of CaF$_2$, wrapped
in a sheet of 25~$\mu$m silver foil, was placed in a Variox cryostat,
and kept at a temperature of 50 K in zero applied magnetic field.  The
Earth's magnetic field was compensated to better than 50~$\mu$T using
active field compensation.
The mean muon lifetime is 2.2~\textmu s, but data can be obtained out to
at least ten times this value at ISIS if collected for several hours.
The asymmetry data were fitted to the function
\begin{equation}
A(t) = A_0 P^\mu(r_{\mathrm{nn}},\lambda;t) + A_{bg},
\label{eq:CaF2fit}
\end{equation}
where $A_0$ accounts for muons which form diamagnetic
states, $A_{bg}$ accounts for muons stopping outside the sample, and
$P^\mu(r_{\mathrm{nn}},\lambda;t)$ is the polarization signal on which we are
focussing. (Here, $r_nn$ corresponds to the nearest-neighbour F--$\mu$ 
distance, and $\lambda$ is the relative adjustment of the nnn
coupling, defined in \eqref{eq:nnnbroadening}.)
The value of $A_0$ is consistent with approximately 35\% of muons
implanting in diamagnetic states, suggesting the remainder are in
muonium states, in agreement with previous work \cite{kiefl}.

\subsection*{DFT$+\mu$ calculations}

The ab initio calculations were performed with the QUANTUM ESPRESSO
package \cite{qe}. The calculations were performed in a supercell containing
$2\times 2\times 2$ conventional unit cells. For the diamagnetic
states considered here, the $+1$ charge state of the muon was
determined by the charge of the supercell.  A muon was placed in
several randomly chosen low-symmetry sites and all ions were allowed
to relax until the forces on all ions and the energy change had fallen
below a convergence threshold.

\subsection*{Calculations of the time evolution of the muon polarization}

The F--$\mu$--F state 
has a time-dependent polarization governed by the magnetic dipolar Hamiltonian
\begin{equation}
\hat{\mathcal{H}} = \sum_{i >j} \frac{\mu_0 \gamma_i \gamma_j}{4 \pi \hbar |\mathbf{r}_{ij}|^3} \Big[ \mathbf{s}_i \mathbf{\cdot} \mathbf{s}_j - 3 \big(\mathbf{s}_i \mathbf{\cdot}\mathbf{\hat{r}}_{ij} \big)\big(\mathbf{s}_j \mathbf{\cdot}\mathbf{\hat{r}}_{ij} \big) \Big],
\label{eq:MagneticDipolarHamiltonian}
\end{equation}\\
where $i$ and $j$ label each nuclear spin and the muon, and $\mathbf{r}_{ij}$ is a vector linking spins $\mathbf{s} _i$ and  $\mathbf{s}_j$, each with gyromagnetic ratios $\gamma_i$ and $\gamma_j$ respectively. For a $\mu$SR experiment undertaken on a polycrystalline sample with no magnetic ordering, and with the instrument in the zero-field (ZF) configuration, a muon enters the sample in a spin-polarized state, with the surrounding atoms in mixed states. Hence, the time evolution of the muon's spin (labelled here as spin $i=0$), $P^\mu(t)$, can be calculated as
\begin{equation}
P^\mu(t) = \frac{1}{2} \Bigg< \mathrm{Tr} \bigg[ \bm{\sigma}_{\hat{n}}^\mu \exp \bigg(\frac{- \mathrm{i}  \hat{\mathcal{H}} t}{\hbar} \bigg) \bm{\sigma}_{\hat{n}}^\mu  \exp\bigg(\frac{\mathrm{i} \hat{\mathcal{H}} t}{\hbar} \bigg) \bigg] \Bigg>_{\hat{n}},
\label{eq:MuonSpinEvolution}
\end{equation}
where $\langle\ldots\rangle_{\hat{n}}$ represents the angular average over $\hat{n}$, and $\bm{\sigma}^\mu_{\hat{n}}$ is the Pauli spin operator for the muon in the direction of $\hat{n}$.

In order to calculate the exponents in \eqref{eq:MuonSpinEvolution},
one needs to diagonalise $\hat{\mathcal{H}}$. Such a matrix has
$2\prod_{i}(2I_i+1)$ rows and columns, (where $I_i$ corresponds to the
spin of the $i$th nucleus and the factor of 2 takes into account the
muon).  The size of the matrices thus grows exponentially with the number of spins being considered. For the simple case of a spin-polarized muon interacting with one fluorine nucleus, $\hat{\mathcal{H}}$ has four eigenstates and three eigenvalues, and the `beats' can be interpreted as the system evolving between such states, as depicted in Fig.~\ref{fig:Fmu}a. 
When more distant nuclei are included in the calculation,
$\hat{\mathcal{H}}$ is split further into more states which leads to
more transitions. 

For CaF$_2$, a direct calculation of equation \eqref{eq:nnnbroadening} results in
$\lambda=0.943$, equivalent to slightly reducing the distance between the nnn 
fluorines and the muon by $5.7\%$ so that they are able to act as a proxy for the 
rest of the lattice. The Hamiltonian can be easily calculated for this system of eleven 
particles (one muon, two nearest neighbour fluorine nuclei and eight next-nearest 
neighbour fluorine nuclei), and has dimension $2048\times2048$, whereas including 
the next shell of fluoride neighbours would become unfeasible for exact diagonalization.
The evaluation of $\sigma_\infty^2$ is performed by calculating terms in the sum out
to some large radius, and then writing all remaining terms out to infinity as an integral.
Our DFT+$\mu$ calculations on CaF$_2$ show that the nnn fluoride ions do move 
towards the muon by a very small distance (approx $0.03$ \AA), and including this
in our calculation of $\lambda$ leads to $\lambda=0.937$.

\section*{Acknowledgements}
J.M.W and S.J.B acknowledge support from EPSRC (grant code EP/N023803/1). Part of this work was carried out at the SFTC-ISIS muon facility, Rutherford Appleton Laboratory, UK and we thank P. Biswas and P. J. Baker for assistance. The entropy calculations were performed on the ARC HPC facility at the University of Oxford.

\section*{Author Contributions}
S.J.B conceived and supervised the project. J.M.W performed the calculations and data analysis. Both S.J.B and J.M.W wrote the paper.

\section*{Competing Interests}
The authors declare no competing interests.

\newpage

\end{document}